\documentclass[prb,
twocolumn,
superscriptaddress,showpacs,amsmath,amssymb]{revtex4}
\usepackage{amsfonts}
\usepackage{bm}
\usepackage{verbatim}
\usepackage{color}
\usepackage{graphicx}

\begin{document}

\title{Flat band and Planckian metal}

\author{G.E.~Volovik}
\affiliation{Low Temperature Laboratory, Aalto University,  P.O. Box 15100, FI-00076 Aalto, Finland}
\affiliation{Landau Institute for Theoretical Physics, acad. Semyonov av., 1a, 142432,
Chernogolovka, Russia}

\date{\today}

\begin{abstract}

We discuss the recent extension of the Sachdev-Ye-Kitaev (SYK) microscopic model \cite{PatelSachdev2019}, which demonstrates the characteristic features of the Khodel-Shaginyan fermion condensate\cite{KhodelShaginyan1990} -- the existence 
of the finite region of momenta, where the energy of electrons is exactly zero (the flat band).  The microscopic derivation of the flat band in this interacting model supports the original idea of Khodel and Shaginyan based on the phenomenological approach. It also suggests that it is the flat band, which is responsible for the linear dependence of resistivity on temperature in "strange metals".

\end{abstract}
\pacs{
}

\maketitle

Recently, in the paper "Theory of the Planck metal", Patel and Sachdev considered a model of  interacting fermions  which describes a Planckian metal at low temperatures, in order to explain the linear
temperature dependence of their resistivity.\cite{PatelSachdev2019}
We show here that the proposed scenario actually describes the formation 
of the Khodel-Shaginyan fermion condensate\cite{KhodelShaginyan1990} (the flat
band). This supports the idea that the 
flat band is responsible for the linear dependence of resistivity on temperature in
"strange metals". 

There are different potential sources of the formation of the electronic flat band with zero energy, see e.g. Ref. \cite{Volovik2018}. In particular it can be formed due to electron-electron interaction. The flat band formed by interaction has been first discussed by Khodel and Shaginyan (KS) in 1990 \cite{KhodelShaginyan1990}, who used the phenomenological Landau theory of Fermi liquid, see also \cite{Volovik1991,Nozieres1992,Volovik1994,HeikkilaVolovik2016} and Fig. 1.
This  dispersionless energy spectrum has a singular density of states. As a result the superconducting gap and  transition
temperature are proportional to the coupling constant instead of the exponential 
suppression in conventional metals with Fermi surfaces.
 For nuclear systems the linear dependence of the superconducting gap on the coupling constant has been
found by Belyaev \cite{Belyaev1961}.
 In a more rigourous manner the flat band induced by interaction has been obtained  in Refs.\cite{Yudin2014,Lee2009}.
Experimentally the merging of levels at the Fermi surface due to interaction has been reported in Refs.
\cite{Dolgopolov2014,Dolgopolov2017} 

In twisted bilayer graphene there is indication that interaction leads to the further flattening of the spectrum
\cite{Marchenko2018,Carr2018} in addition to the geometrical/topological flattening caused by the magic angle  twist \cite{Cao2018a,Cao2018b}.

\begin{figure}
\centerline{\includegraphics[width=\linewidth]{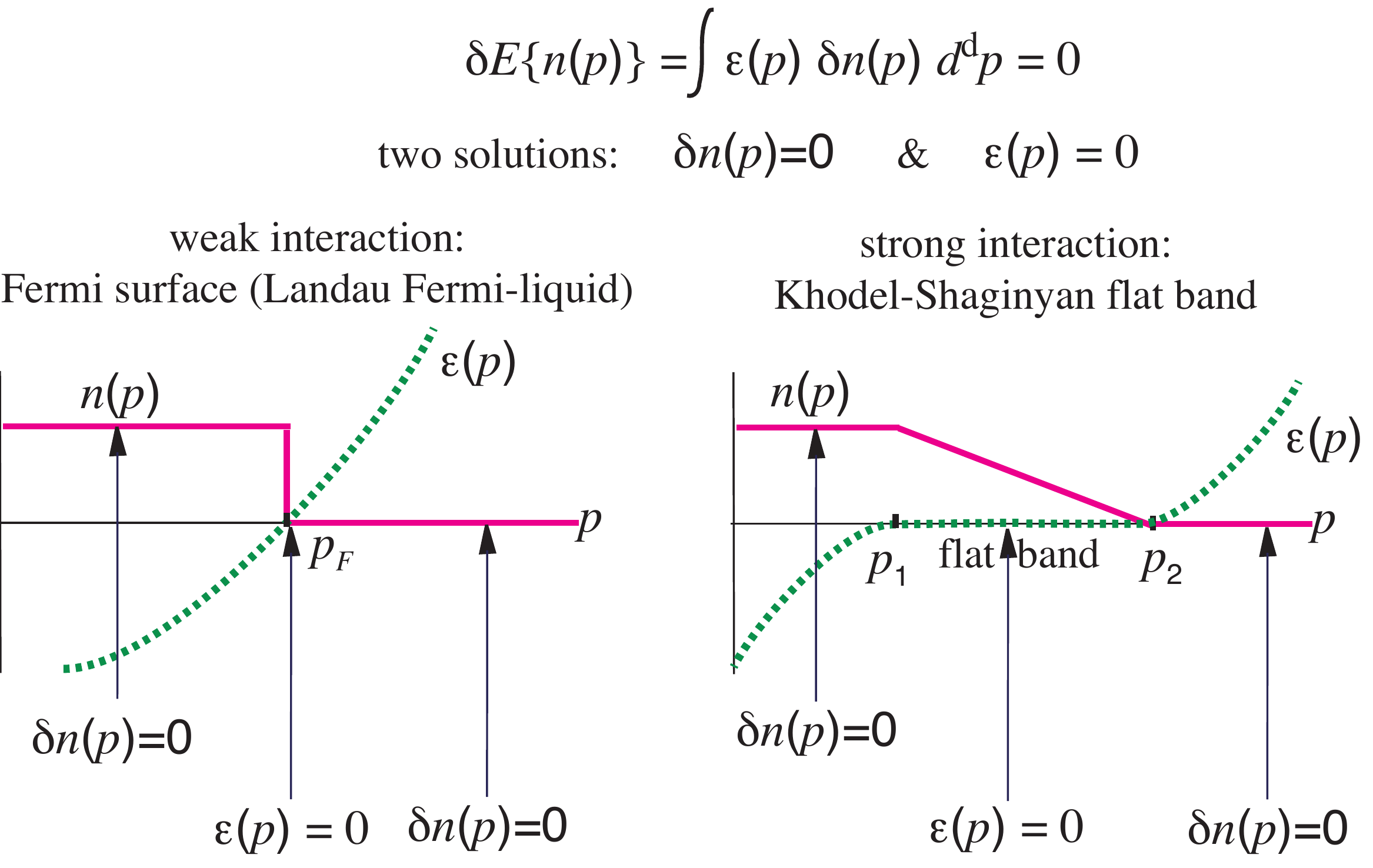}}
\label{KS} 
  \caption{Consequences of Landau theory of Fermi liquid. Variation of the energy functional over the occupancy $n({\bf p})$ gives two possible solutions: $\epsilon({\bf p})=0$ and $\delta n({\bf p})=0$ ($n({\bf p})=0$  or $n({\bf p})=1$). 
 {\it left}: The Landau Fermi liquid, where the solution $\epsilon({\bf p})=0$ takes place on Fermi surface.
Outside of Fermi surface one has either $n({\bf p})=0$  or $n({\bf p})=1$. {\it right}: Khodel-Shaginyan flat band, where the solution $\epsilon({\bf p})=0$ takes place in the finite region of momentum space.  In this region $0<n({\bf p})<1$.
}
\end{figure}

In recent paper by Patel and Sachdev\cite{PatelSachdev2019} the lattice extension  of the Sachdev-Ye-Kitaev (SYK) model has been used to study the problem of the "bad metal"   with the universal linear dependence of resistivity on temperature \cite{Legros2018,Nakajima2019,Cao2019,Brown2019}. However, it appears that 
signatures of the KS flat band in  Figure 1  ({\it right panel}) are very similar to those  in Figs. 2a and 3a from  Patel and Sachdev (PS)\cite{PatelSachdev2019}. Indeed, Fig. 2a from the PS paper shows the occupancy $n({\bf p})$, which exhibits  the same behavior as $n({\bf p})$ in Fig. 1 ({\it right panel}), with the finite region  where $0<n({\bf p})<1$. According to Khodel-Shaginyan, in this region the quasiparticle energy should be zero. And this is clearly seen from the electron spectral density shown  in Fig. 3a of the PS paper.
So one may conclude that the extended SYK model provides another possible realization of the KS flat band.

 That is why the extended SYK model can be used for studying different properties of the materials which experience formation of the KS flat band, including possibly the  "bad metal" behavior. 
In this model, the universal linear dependence of resistivity on temperature has been obtained  \cite{PatelSachdev2019} in the regime, where the signatures of the flat band are transparent. From that one may conclude that the phenomenon of Planck metal or bad metal is the consequence of the Khodel-Shaginyan flat band emerging in this model.
The idea that the flat band may serve as the origin of the "strange metal" behavior has been suggested earlier, see e.g. Ref.  \cite{Shaginyan2013}, and recent papers \cite{Khodel2018,Shaginyan2019}.

This work has been supported by the European Research Council (ERC) under the European Union's Horizon 2020 research and innovation programme (Grant Agreement No. 694248).


\end{document}